\documentclass[conference]{IEEEtran}
\IEEEoverridecommandlockouts
\usepackage{cite}
\usepackage{amsmath,amssymb,amsfonts}
\usepackage{algorithmic}
\usepackage{latexsym}
\usepackage{graphicx}
\usepackage{amsfonts,amssymb,amsmath}
\usepackage{hyperref}
\usepackage{multirow}
\usepackage{color,soul}
\usepackage{caption}
\usepackage{subcaption}
\usepackage{graphicx}
\usepackage{listings}
\usepackage{enumitem}

\def\BibTeX{{\rm B\kern-.05em{\sc i\kern-.025em b}\kern-.08em
    T\kern-.1667em\lower.7ex\hbox{E}\kern-.125emX}}
\begin{document}

\title{Engineering an IoT-Edge-Cloud Computing System Architecture: Lessons Learnt from An Undergraduate Lab Course}

\author{
    \IEEEauthorblockN{Jasenka Dizdarevi{\'{c}} and Admela Jukan}
    \IEEEauthorblockA{Technische Universit\"at Braunschweig, Germany}
   \IEEEauthorblockA{\{j.dizdarevic, a.jukan\}@tu-bs.de}
}

%author{\IEEEauthorblockN{1\textsuperscript{st} asenka Dizdarevi{\'{c}}}
%\IEEEauthorblockA{\textit{Technische Universit\"at Carolo-Wilhelmina zu Braunschweig} \\
%Braunschweig, Germany  \\
%j.dizdarevic@tu-bs.de}
%\and
%%\IEEEauthorblockN{2\textsuperscript{nd} Admela Jukan}
%\IEEEauthorblockA{\textit{Technische Universit\"at Carolo-Wilhelmina zu Braunschweig} \\
%Braunschweig, Germany \\
%a.jukan@tu-bs.de}

\maketitle

\begin{abstract}
With the rapid advances in IoT, edge and cloud computing solutions, it is critical to educate and train students in computer science and engineering in various aspects of IoT-edge-cloud (IoT-E-C) system architecture implementations. We outline the design and 
development of an undergraduate laboratory course that sets the goal of implementing various interfaces and communication protocols to connect IoT, edge and cloud 
computing systems and evaluating their performance. The lab setup is modular and based on open source tools. In the IoT context, it consists of low-cost processing platforms with various sensors and actuators. In the edge and cloud computing context, we implement 
and deploy single board computers and Firebase cloud solutions, respectively. The modular lab setup allows students  to engineer and integrate various communication protocol solutions, including MQTT, COAP and HTTP. In addition to the  system implementation, students can evaluate and benchmark the performance of the entire system.  
\end{abstract}

\begin{IEEEkeywords}
    Edge and cloud computing, communication protocols, IoT
\end{IEEEkeywords}

\section{Introduction} \label{intro}

The unstoppable trend towards the combined IoT, edge and cloud computing systems has lead to an increased demand for educated workforce in corresponding areas of computer science and engineering. This has furthermore driven efforts in creating computer science and computer engineering courses that can implement and integrate IoT devices in cloud and edge computing systems \cite{He, Burd, Lee, 8340468}. Currently available courses reflect the ongoing quests towards the specific focus of the course, such as testing different hardware boards, or developing domain applications with IoT devices, such as in healthcare applications \cite{Farhat}. Today, in addition to a number of cloud and edge based IoT commercial platforms, numerous open-source solutions can be used to further broaden the participation in such courses \cite{Guth2018}. Since the technology choice is both broad and diverse, and the technology trends and the contents covered are constantly evolving, scoping such courses towards specific goals is rather critical.  

We propose and outline the design and development of an undergraduate laboratory course, referred to as Network-of-Things Engineering Lab (NoteLab), - that sets the goal of implementing various interfaces and communication protocols to connect IoT, edge and cloud computing systems and evaluating their performance. Unlike other courses that usually cover individual and separate areas of either IoT, or edge, or cloud computing subsystems, our course offers the implementation of the entire system, and the subsequent evaluation and benchmarking of the end-to-end system performance. The lab setup is highly-modular and based on open source tools. It includes three contexts: IoT, edge and cloud, as a model for separation of concerns implying that any device can be deployed in a specific context.  This is in contrast to a commonly used computing hierarchical architecture, with cloud computing at the top of hierarchy \cite{7868354}. The modular lab setup allows students  to engineer and integrate various communication protocol and interface solutions, including MQTT, COAP and HTTP. 
%In the edge and cloud system context, we implement  and deploy single board computers and Firebase cloud solutions, respectively. 
We define the following main learning objectives in the course:
\begin{itemize}[noitemsep,topsep=0pt]
\item{O1. Engineering the interfaces and the related communication protocols in an IoT embedded system setup, as an integral part of the integrated IoT-edge-cloud computing architecture;}
\item {O2. Integrating a diverse set of the specific hardware solutions in IoT and edge computing context (e.g., sensors, actuators, single-board-computers, and microcontroller platforms);}
\item{O3. Applying open-source software solutions for resource-constrained embedded devices (including Arduino IDE sketches, ARM compatible docker images, mqtt proxy, coap proxy and http proxy python scripts);}
\item {O4. Executing and testing sample sensing and actuation applications (e.g., temperature, humidity and motion sensing);}
\item{O5. Measuring and benchmarking system performance (e.g., latency and power consumption).}
\end{itemize}

The rest of the paper is organized as follows. Section \ref{background} presents the related work. Section \ref{arch} describes the NoteLab architecture design and the learning scope. Section \ref{setup} presents the details behind the individual task units in the lab, along with a discussion on lessons learnt. Section \ref{conclusion} provides conclusion and outlook.

\section{Related work and our contribution}\label{background}

The driving force behind the innovation and implementation of IoT system solutions in recent years, which has also led to an increase in the related electrical and computer engineering courses, is the availability of low-cost, yet highly performant IoT devices \cite{He}. Many hands-on courses developed to date focus on IoT laboratories \cite{karvinen2018iot}. Paper \cite{Burd} presents an extensive IoT courses survey defining different categories, ranging from introductory IoT courses, more advanced IoT certification multi-courses, over to most common approach of studying and developing domain-specific IoT applications, such as in agriculture, transportation or healthcare. Paper \cite{Kurkovsky} examines different hardware platforms for IoT-centric courses, and emphasizes the challenges that the course instructors need to pay attention to when choosing specific hardware and software solutions. Papers \cite{barendtnew} and \cite{Nelke} describe various approaches to teaching new IoT concepts through elective courses. 

Paper \cite{8340468} presents a newly developed embedded system IoT course, explaining to students and their instructors specific aspects of data processing in IoT systems. As the data processing has shifted more towards cloud computing context, major efforts in computer engineering also shifted focus to only collecting data with IoT devices, and their processing in the cloud \cite{Maitra}. More recently, data processing is done in edge and fog computing \cite{AI201877}, which has also brought a notable interest in combined IoT-edge-cloud solutions, both in industry and academia \cite{8089336, Mahmud}. Paper \cite{Grammenos} presents a system oriented graduate course, covering the concepts of edge and cloud computing in integrated IoT systems. This more system oriented direction has successfully integrated aspects of networking and communication protocols, even in undergraduate IoT lab courses, such as in \cite{Lee}. 

In terms of our specific novel contributions, we greatly benefitted from previous work, especially in our initial choices of hardware platforms, such as from \cite{Kurkovsky, 8340468}. We adopted similar goals set in some courses, such as to creating an industrially relevant IoT system in \cite{Grammenos}. We took  a different approach, however, and worked with open source and non-proprietary hardware and software, with the goal of enabling easy reproducibility of the course, albeit possibly of a lesser industrial interest. Also notable is related work \cite{Lee}, addressing not only the integrated IoT-edge-cloud computing but also introducing different IoT protocol solutions. This course however uses a different methodology and software systems. In our approach, we do not use stand-alone lab units, but design all units as inter-related and building upon each other, where students can gain significant problem-solving skills as they are trying to plan and connect individual tasks. In addition, we use containerization, including Docker, but also as applied to Kafka, Firebase and Mosquitto, which provides additional training in software engineering. Our paper does not include a detailed taxonomy based evaluation of student learning outcomes \cite{chan2002applying}, as the focus has been mainly on technical aspects of the course implementation. In future work we plan an extension that will include this kind of evaluation, as it will help improve our learning framework and outcomes.

\section{NoteLab architecture design}\label{arch}

\begin{figure*}[ht]
  \centering
  \includegraphics[width=0.90\textwidth]{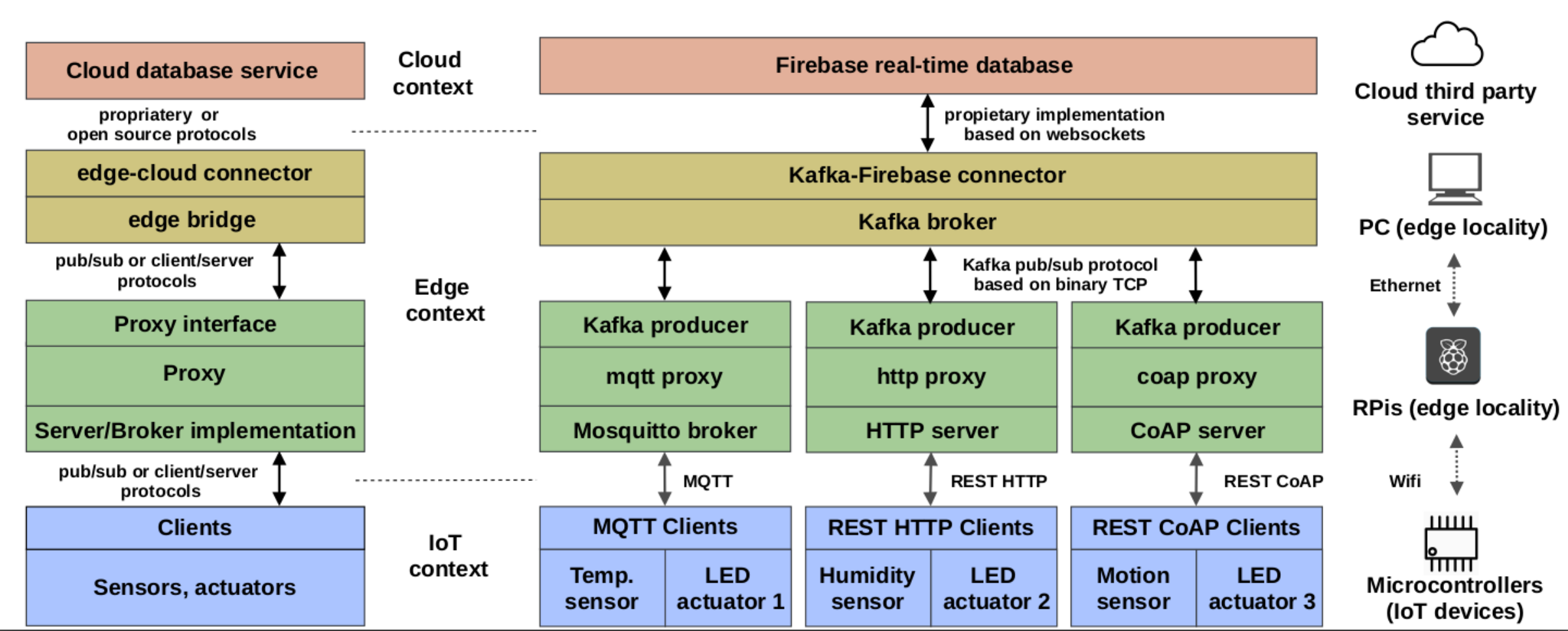}
\caption{NoteLab reference architecture: a) high-level overview and b) implementation}\label{fig:arch}
\vspace{-0.4cm}
\end{figure*}  

\subsection{Reference architecture} \label{approach}

We start with the envisioned system architecture shown in Fig.~\ref{fig:arch}, representing both the high-level architectural concept as well the implemented system solution. We divide the architecture into various \emph{context}: IoT-, edge computing- and cloud computing context. In our approach, we found it useful to refer to a \emph{context} as a separation of concerns, as opposed to \emph{layers} that often imply hierarchy. In other words, there is no assumption on any specific hierarchy in the architecture, and computing and data processing can happen in all three context, and on any device put in context.  

In the \emph{IoT context,} we envision resource constrained devices and low-cost processing platforms, including various sensors and actuators. These sensors and actuators are the principal generators of data in the architecture and will have a client role (Fig.~\ref{fig:arch}a). The way how these clients are going to send the data to other devices depends on the implementation, assuming that both the communication protocols that are based on publish/subscribe and client/server (also known as request/reply) interaction model are potential choices.

In the \emph{edge computing context,} we define two types of devices, one with the role of connecting the edge with the cloud database service (edge-cloud connector in Fig.~\ref{fig:arch}a), for which in the implementation we will use an already developed open source solution; and the other with the role of a proxy, which will be developed by the students with the intent of forwarding data from the devices in IoT context to the connector. In order to establish communication with IoT devices, the proxy solution is going to include a server or a broker implementation (depending on the interaction model of the communication protocol) as its southbound interface. It will also include a northbound interface for forwarding data to the edge-cloud connector. For the latter communication, as it is based on an open-source pre-developed solution, a publish/subscribe based protocol is used. 

The third context is the \emph{cloud computing context}, where we envisioned a simple cloud database service for client's data storage and synchronization purposes. The communication between the database and the edge-cloud connector will be maintained through proprietary or open source protocols, depending on the implementation choice of the cloud service.

The architectural representation containing implementation details is shown in Fig.~\ref{fig:arch}b, illustrating various devices put in IoT, edge or cloud context. Here, for the IoT context implementation, we use six microcontroller based devices, with the corresponding sensors and actuators, including  temperature, humidity, and motion sensors as well as LED actuators. 
Localized in the edge context, we differentiate between two types of devices: single board computers (Raspberry Pi, RPi) and off-the-shelf desktop computers. 

As a part of the edge context implementation Fig.~\ref{fig:arch}b illustrates three devices implemented with single board computers, which are used for our proxy solutions (mqtt, http and coap proxy), each running the corresponding software. The connection between the microcontrollers and single board computers is established through their WiFi interfaces. The other type of edge context device, implemented on the desktop computer is used for the edge-cloud connector implementation, shown as \emph{Kafka-Firebase connector} in Fig.~\ref{fig:arch}b, an open source solution available from github \cite{kafkagit}. The connection between two types of edge context devices is through a switch over their ethernet interfaces. This particular edge-cloud connector implementation was selected due to our choice of the Firebase as the cloud database service for NoteLab.

And finally, for the cloud context implementation, as mentioned, we deploy Google's Firebase cloud solution \cite{firebase}. The Firebase was selected due to being an open source and its ability of storing the client's data locally through the edge context's connector, and later automatically synchronizing with the cloud \cite{carpio2020engineering}. In addition, setting up a Firebase Database is easy and takes only few steps, providing quicker learning curve for students. 

The setup and implementation of a few distinct interface and communication protocol solutions to establish communication between devices used in various contexts is one of the salient features of NoteLab. Current state of the art points at MQTT (Message Queue Telemetry Transport) as communication protocol of choice due to its maturity and performance. This is followed closely by HTTP (HyperText Transfer Protocol) as the widely adopted solution and a second best choice of developers \cite{EclipseIoTWorkingGroup2018}. Finally,  CoAP (Constrained Application Protocol) is a well-known IoT messaging standard communication alternative to these two protocols \cite{Bhattacharjya2020} due to its lightweight characteristics and comparatively better performances in resource-constrained environments \cite{Dizdarevic}.
In the implementation part of architecture shown in Fig.~\ref{fig:arch}b, we illustrate the choice of interfaces at the context boundaries based on few possible choices for related communication protocols: MQTT, HTTP and CoAP.
When using publish-subscribe MQTT communication protocol, at the boundaries between IoT and edge context, RPi edge localized device interfaces the lower level devices of the IoT context (microcontroller boards with their attached sensors and actuators) with a MQTT Mosquitto broker. These devices serve as MQTT publisher and subscriber clients. Since the architecture is extensible, other protocols can be modularly implemented in parallel, where edge computing context would implement the server (HTTP or CoAP) while the IoT context would implement corresponding interfaces for client applications, based on REST standard software architecture.

In the remaining parts of the reference architecture, the interface between the Kafka based edge-cloud connector and Firebase database is based on a Firebase proprietary communication implementation over websockets. In the high-level overview we saw that the communication protocols and interfaces between proxy solution and edge-cloud connector are based on publish/subscribe paradigm. In the implementation this is achieved with interfacing the Kafka based connector (comprised of multiple components, one being a Kafka broker) with Kafka producer as the northbound interface on the implemented proxy solutions. The communication protocol used here is Kafka's native binary TCP protocol, an integral part of Kafka implementation which as such is not required for students to understand all the details. Instead, students are instructed that when the software tools require native protocols to be used, it is necessary to develop the corresponding \emph{proxies}, as described in detail in the following sections.

%%%%%%%%%%%%%%%
\subsection{Scoping the learning framework}\label{framework}

To appreciate the broadness of the subject matter, we now briefly give an overview of the learning scope, and outline the reasons behind our choices made. Table~\ref{table:hwtable} gives an overview of the hardware development kit, with device-corresponding operating systems or firmware within a specific context as previously described. 

\begin{table}[h!]
  \caption{NoteLab hardware with corresponding operating systems}
  \setlength{\tabcolsep}{3pt}
  \begin{tabular}{|p{34pt}|p{138pt}|p{61pt}|}
\hline
\vfill Context &\vfill Hardware & Operating system or firmware \\ 
\hline
\vfill \textbf{IoT}& Single board ESP8266 microcontrollers: NodeMCU and WeMos D1 R2& optional-NodeMCU firmware \\
\cline{2-3}
& Sensors: DHT11 temperature and humidity sensor, HC-SR04 ultrasonic sensor, Passive Infrared Sensor, LED actuators, set of resistors and jumper wires & \vfill / \\
\hline
\textbf{Edge}&Single board computers: RPi 3/4 Model B &Raspberry Pi OS\\
\cline{2-3}
&Desktop computer & Ubuntu 20.04 \\
\hline
\textbf{Cloud} & \vfill Desktop computer (Firebase Console) & \vfill Ubuntu 20.04 \\
\hline
\end{tabular}
\label{table:hwtable}
\vspace{-0.3cm}
\end{table}

The choice of IoT hardware platforms is based on how widely available and used they are, including the availability of adequate tutorials and developer's support \cite{Kumar, 7474533}. In the IoT context, we chose two types of microcontroller development boards \cite{esp8266documentation}, i.e., NodeMCU and WeMos D1 R2. The practical reasons behind this choice is their WiFi connectivity  support (as they are based on ESP8266) critical to building a networked system of IoT devices. In addition, these microcontrollers are programmable using Arduino IDE, a user-friendly environment as it comes with an abundance of online  programming examples freely available to students. Finally, numerous low-cost sensors and actuators that were initially manufactured for the Arduino platform are also compatible with these microcontrollers. In NoteLab we use DHT11 temperature and humidity sensor, HC-SR04 ultrasonic sensor, Passive Infrared Sensor (PIR) motion sensor, LED actuators, along with a set of jumper wires and resistors for connecting the circuits. For microcontroller configurations, the course manual includes instructions on how to configure the relevant ESP8266 board parameters in Arduino IDE. This way, students can easily use the code programmed on microcontrollers.

In the edge context, as mentioned we use two types of devices, i.e., single board computer and off-the-shelf desktop computer. For single board computers, each student is provided with either a Raspberry Pi 3 Model B+ or Raspberry Pi 4 Model B, with their corresponding operating systems pre-installed. RPis are generally very popular among students, due to being low-cost, with powerful computing and interfacing features, both in hardware and software. In NoteLab, students use RPis both in the IoT context as simple workstations to program the ESP8266 boards and as a computing resource for receiving and processing sensor data. The off-the-shelf desktop computer serves as the edge-cloud connector. Its southbound interface (Kafka's broker) communicates with RPis, while its northbound interface (Firebase interface) connects to the cloud real-time database through a Firebase Console. The OS for the desktop computer is always chosen as open-source, in our case the latest Ubuntu version.

In the cloud context, students are expected to create a real-time database through Firebase console and to generate an authorization key to connect the said database to edge-cloud connector. In other words, the usage of the cloud is scoped for data storage purposes only.

A more detailed overview of the software development solutions used to scoping the course is outlined in Table~\ref{table:swtable}, with MQTT as the protocol of choice. (Other protocols would result in different table entries and are not listed here for brevity). It is first important to consider that also students without programming skills are able to learn the basics of different programming languages, different software solutions, and finally, an increasingly important concepts of virtualization and containerization techniques. The importance of introducing these concepts, which have long been the critical skills in high-tech industry, cannot be overstated \cite{Boettiger,skills}. 

When MQTT protocol is used, different types of sensors and actuators are implemented using Arduino IDE  for programming MQTT clients. Programming MQTT in Arduino IDE is well documented by numerous tutorials and research activities and hence easy to use \cite{Keophilavong2019, 10.1145/3323503.3349546,8660936, 8441423}. In NoteLab, students are given specific instructions on how to create the so-called Arduino sketch codes (based on C/C++) for MQTT publisher and MQTT subscriber. (The examples of this sketch code will be given in the following section.) While this code is based on C/C++, for programming in Arduino IDE, students are not required to actually know C or C++ programming. In other words, for developing simple sketches to be used in NoteLab, students are given specific scripts and instructions. %Finally, since the programming at this level is actually done on microcontrollers, which are low-level programmable devices, the docker containerization in this layer is not applicable.

In terms of virtualization techniques, we choose Docker containerization \cite{rad2017introduction}. One of the known benefits of Docker-based containerizations is its performance in CPU and memory utilization, which students can experience as especially relevant in combination with resource constrained IoT devices \cite{Bellavista2017, 9037775}. We include in Table ~\ref{table:swtable} a separate column that shows which system components in the reference architecture use containerization. To acquire \emph{dockerization} skills, students use two pre-installed docker images in the edge computing context. One image is used to implement a Mosquitto MQTT broker (developed in C, \cite{eclipse}) on RPi devices, while the other one to implement \emph{Kafka-Firebase connector} (developed in Java) on the desktop computer. Since these images will be used as pre-installed solutions, students only focus on  modifying the related configuration parameters. On the other hand, students are requested to actually develop a \emph{mqtt proxy}, albeit based on detailed instructions provided to them. This is critical, since the objective is to establish communication between the two docker based system edge components, i.e., Mosquitto broker and Kafka based edge-cloud connector. To program this application students are requested to use a Python script, and dockerize it. 

\begin{table}[h!]
  \caption{NoteLab system components with MQTT as protocol of choice}
    \setlength{\tabcolsep}{3pt}
    \begin{tabular}{|p{27pt}|p{44pt}|p{68pt}|p{48pt}|p{35pt}|}
  \hline
 Context &Device & System component (software solution) & Programming language & Dockerized\\ 
\hline
\textbf{\vfill IoT}& Sensor attached to ESP8266 board &\vfill Client (MQTT publisher) & Arduino sketch (code unit based on C/C++)& \vfill No\\
\cline{2-5}
& Actuator attached to ESP8266 board & \vfill Client (MQTT subscriber) & Arduino sketch  (code unit based on C/C++) & \vfill No\\
\hline
\textbf{\vfill Edge}& RPi 3/4 Model B &Broker (Mosquitto) & C & Yes\\
\cline{3-5}
&  & mqtt proxy & Python script & Yes\\ 
\cline{2-5}
&\vfill Desktop computer & Edge-cloud connector (\emph{Kafka-Firebase connector}) & \vfill Java program & \vfill Yes\\
\hline
\end{tabular}
\begin{tabular}{|p{27pt}|p{213pt}|}
  \textbf{Cloud} & Firebase real-time database \\
  \hline
\end{tabular}
\label{table:swtable}
\vspace{-0.4cm}
\end{table}

Summarized over Tables~\ref{table:hwtable} and ~\ref{table:swtable}, we design the NoteLab to include the following hardware and software systems, i.e., 

\textbf{Hardware systems}
\begin{itemize}
  \item  ESP8266 complete development boards (Arduino compatible e.g. NodeMCU or WeMos D1 R2)
  \item  Arduino sensor and actuator kit  
  \item  Micro USB connector (uploading the code, update firmware, charging the battery)
  \item  Breadboard, Jumper wires and Resistor Kit 
  \item  Raspberry Pi single-board computers
  \item  Off-the-shelf desktop computer
  \item  WiFi router
 \end{itemize}
 
 \textbf{Software systems}
 \begin{itemize}
   \item  Arduino IDE with added support for ESP8266 boards 
   \item  Arduino IDE libraries for WiFi support, communication protocols and different sensors(e.g. DHT11 library for temperature and humidity sensor; PubSubClient library for MQTT)
   \item Docker for the ARM architecture
   \item Docker and docker-compose for the x86 architecture
   \item Python programming language.
  \end{itemize}

\section{Laboratory Setup and Task Units in Context}\label{setup}

Before going into details of each task unit, a summarized mapping of task units, learning objectives, student's learning outcomes and NoteLab architecture is outlined in Table~\ref{table:outcomes}. 

\begin{table}[h!]
  \caption{Mapping of task units with learning outcomes, learning objectives and the reference architecture}
  \setlength{\tabcolsep}{3pt}
  \begin{tabular}{|p{14pt}|p{34pt}|p{123pt}|p{55pt}|}
  \hline
  \textbf{Task unit} & \textbf{Learning objectives} & \textbf{Learning outcome} & \textbf{Computing context}\\ 
\hline
\centering 1 & \centering O2 & Configure a microcontroller & IoT  \\
\hline
\vfill \centering 2 & \vfill \centering O2 & Connect sensors and actuators based on circuit layouts  & IoT  \\ 
\hline
\vfill \centering 3 & \centering O3, O4 & Demonstrate programming of three sensors with Arduino IDE sketches & IoT  \\ 
\hline
\vfill \centering 4 &\vfill \centering O3, O4 &  Setup WiFi network using RPi as an access  point  & IoT and edge  \\ 
\hline
\vfill \centering 5 & \vfill \centering O3 & Use containerization tool to run a proxy solution on Raspberry Pi & edge \\ 
\hline
\vfill \centering 6 & \vfill \centering O1 & Establish communication between IoT devices and edge localized proxy & IoT-edge interface\\ 
\hline
 \centering 7 &  \centering O5  & Statistical performance analysis & IoT and edge \\ 
\hline
\centering 8 & \centering O3  & Configure a cloud database service & cloud \\ 
\hline
\vfill \centering 9 & \vfill \centering O1, O3 & Configure and run a containerized edge-cloud connector & IoT-edge-cloud interfaces\\ 
\hline
\vfill \centering 10 &\vfill \centering O5  & Statistical performance analysis of the entire system &  IoT, edge and cloud  \\ 
\hline
\end{tabular}
\label{table:outcomes}
\vspace{-0.2cm}
\end{table}

\vspace{-0.2cm}

\subsection{IoT Context (Task Units 1-4)}\label{iotcontext}

\begin{figure}[h]
    \centering
     \includegraphics[width=0.95\columnwidth]{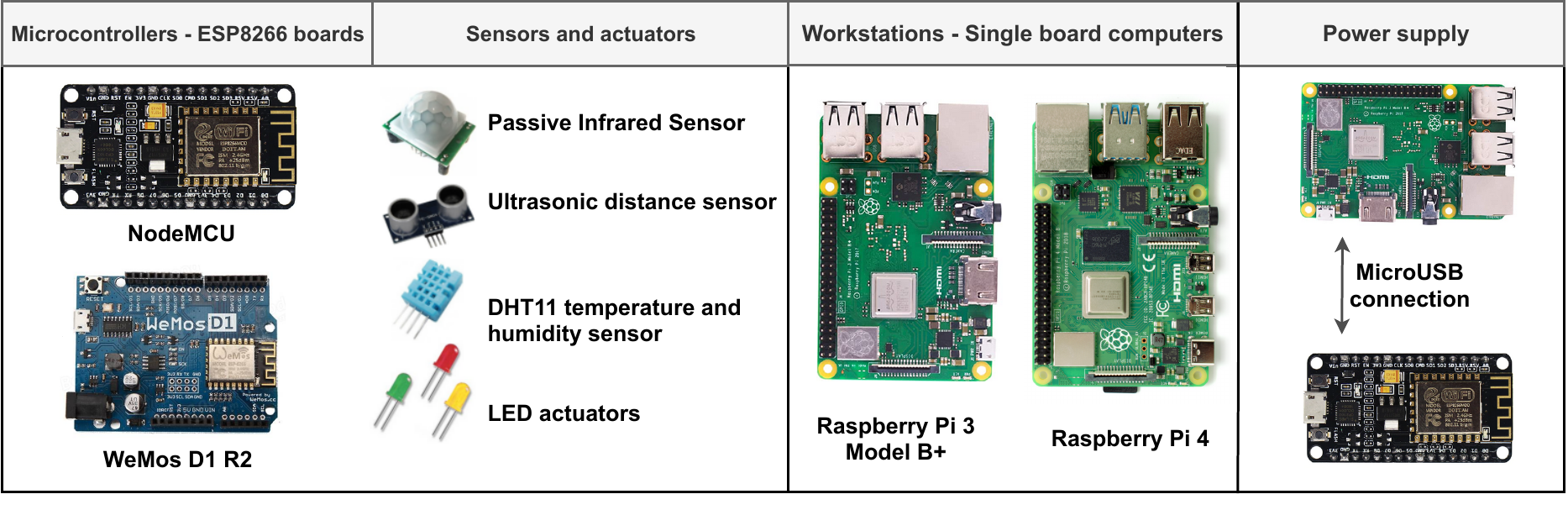}
      \caption{\label{fig:iotkit} Basic kit in IoT context}
      \vspace{-0.2cm}
  \end{figure}

The familiarization with IoT devices is the foundation of this course. In IoT context, students are learning to configure microcontrollers and sensors, to wire the IoT devices with the devices in  edge computing context and to writing and running the software that collects the measurements. Here, each student receives two microcontrollers and one single board computer (RPi). In addition, each student receives a set of three types of sensors and three LED actuators, with jumper wires and set of resistors that will help them connect the circuits. This is shown in the left part of Fig.~\ref{fig:iotkit}, and it represents the basic kit for developing an IoT device with sensing and actuating functions. In this context also single board computers are used, serving as workstations to connect the microcontrollers via MicroUSB cable for energy supply (i.e., battery-less operation). 

\noindent\textbf{Task Unit 1 (IoT device setup):} \label{task1-1}
The first task unit starts with detailed instructions and all necessary information about microcontrollers, sensors and actuators, based on the corresponding survey of their specifications, supported interfaces, circuit physical layout, pin definition, etc. After connecting the microcontrollers to single board computers, students are required to actually use the single board computer's \emph{commandline} (the instructions will depend on the OS running on the device, pre-installed by instructors) to configure microcontrollers by flashing the NodeMCU firmware. (It should be noted that this step is not necessary for programming in Arduino IDE). The version of the firmware used is to be pre-downloaded on RPis and it can be download from \cite{nodemcu} under a name such as  ${\textstyle nodemcu-master-7-modules-x-float.bin}$. Another requirement is to install the firmware flashing tool, called \emph{esptool}. The  \emph{esptool} installation and utilization instructions from \emph{commandline} are illustrated in the code block below. Independently of the board used, all microcontrollers need to be manually flashed with NodeMCU firmware. This enables the students to find out which serial port of RPi connects to the microcontroller. Using that serial port (for example \emph{ttyUSB0}) it is possible to flash the firmware. The code block consists of following commands: 

\lstset{numberstyle=\tiny, morecomment=[l]{//}, commentstyle=\color[rgb]{0,0,0.70}\ttfamily\itshape\footnotesize}  
\begin{lstlisting}[linewidth=\columnwidth,breaklines=true,basicstyle=\footnotesize]
  $ sudo apt-get install esptool
  $ dmesg  //Output:
  [15705.320141] usb 1-1: cp210x converter now 
  attached to ttyUSB0
  $ sudo esptool --port /dev/ttyUSB0 
  write_flash 0 nodemcu-master-7-modules-x-float.bin
  \end{lstlisting}

  \vspace{-0.3cm}

\begin{figure}[h]
   \centering
    \includegraphics[width=0.94\columnwidth]{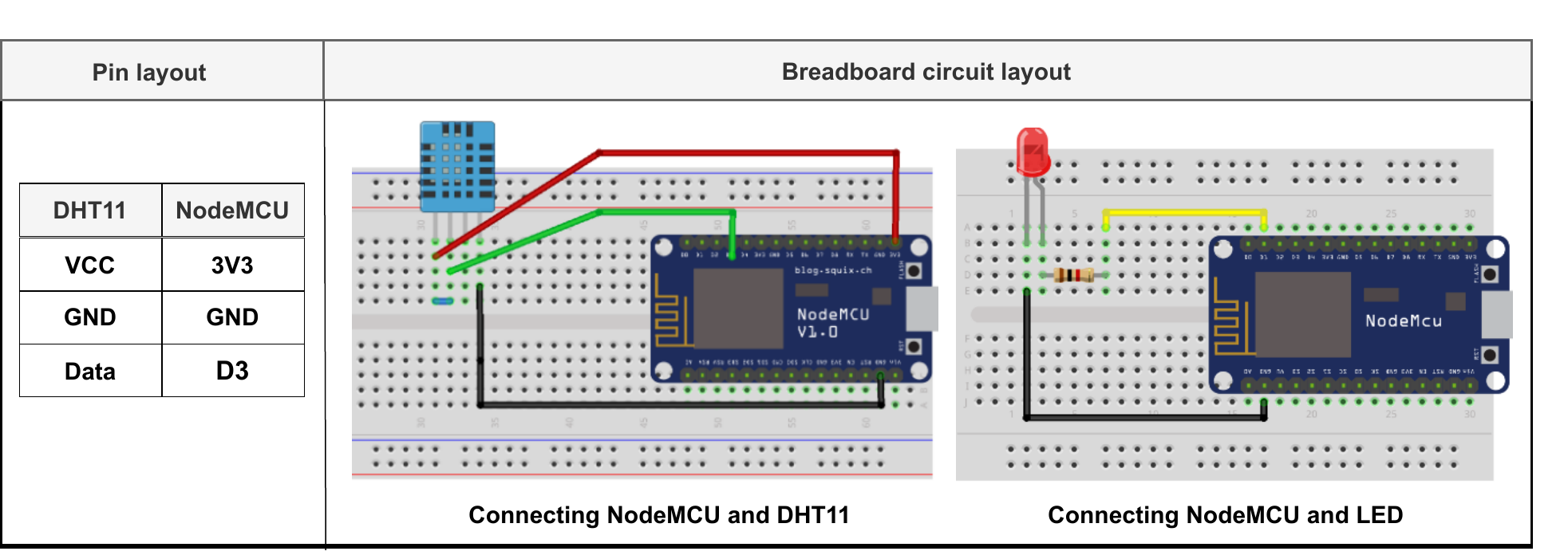}
     \caption{\label{fig:circuit} Interfacing microcontrollers with sensors and actuators}
     \vspace{-0.2cm}
\end{figure}

\noindent\textbf{Task Unit 2 (Circuit setup):} \label{task1-2} In this task unit, students learn how to connect each of the sensors and actuators to the microcontroller. Students receive a breadboard circuit layout and a pin layout. This is illustrated in Fig. \ref {fig:circuit} with LED actuator and DHT11 temperature and humidity sensors connected to NodeMCU board. Based on the circuit scheme and pin layout table, microcontrollers, sensors and actuators are placed on their corresponding breadboards and connected with jumper wires. The instructions provided to students include information about which type of resistor is needed for component protection. It should be noted that the circuits are different for different boards. In the example shown, LEDs have two pin interfaces, one with a long leg (anode) for positive supply that is connected with a yellow wire to one of the GPIO (General Purpose Input/Output) pins of NodeMCU, and which is then used to carry digital or analog signals. The smaller leg (cathode) is used for negative supply that will be connected with a black wire to the ground pin (GND) of a microcontroller. The resistor of value 200 $\Omega$ is added in series with the LED. In other example, when connecting NodeMCU with DHT11, three of the pins on each side are to be connected with jumper wires: DHT11's pin that supplies power for the sensor - VCC pin connects to +3.3 V pin of NodeMCU (red wire) and DHT11's ground pin to the ground pin of the NodeMCU (black wire). The Data pin on the DHT11 sensor connects to one of the GPIO pins of NodeMCU (here, pin D3) with a green jumper wire. Finally, a 10 k$\Omega$ resistor is added between VCC and Data pin of DHT11, as can be seen in Fig.~\ref{fig:circuit}. This task unit needs to be repeated for all sensors. 

\vspace{-0.2cm}

\begin{figure*}[h]
  \centering
  \includegraphics[width=0.92\textwidth]{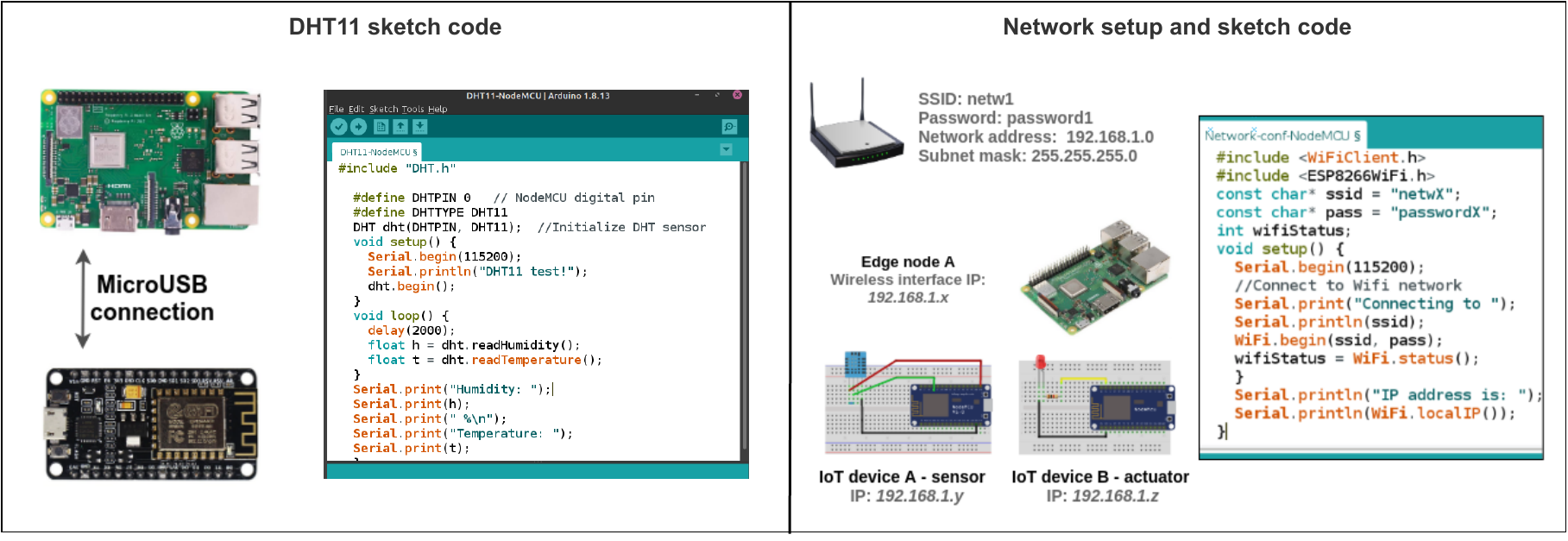}
\caption{IoT context - Programming microcontrollers with Arduino IDE}\label{fig:iotarduino}
\vspace{-0.4cm}
\end{figure*}  

\noindent\textbf{Task Unit 3: (Programming microcontrollers)} In this task unit, students are introduced to Arduino IDE and to programming microcontrollers through writing, compiling, and uploading the code.  First, students install the latest version of Arduino IDE on the RPi used to configure ESP8266 boards. To test how DHT11 sensor can be used, students test Arduino IDE sketch code provided to them, as illustrated in Fig.~\ref{fig:iotarduino}. To upload the code, however, students are requested to find under \emph{Tools} an option to manage libraries in Arduino IDE. This is necessary to making sure that  \emph{DHT.h} library, which is the corresponding library for sensor exemplified here, was installed. To read the output of Arduino IDE, - in this case temperature and humidity values, students use \emph{Serial Monitor} - a separate pop-up window from Arduino IDE that acts as terminal, also to be found under \emph{Tools}. This \emph{Serial Monitor} is used for IoT setup verifications, since in case that the circuit components have not been correctly connected, the expected outputs after compiling and uploading the code will not be shown. In case there were no errors, the output on the monitor will show sensor readings, which for DHT11 are the values measured on temperature and humidity sensors. Arduino IDE sketch code is also tested for LED, which is commonly used as a first tutorial in Arduino IDE, since it does not require inclusion of libraries. This sketch code is to be uploaded to the second of the two microcontrollers given to students. %For the remaining sensors, students will not be provided the entire code, instead they will be instructed about which set of libraries to include and how to program the sensors by themselves. For successfully completing the course, students have to be able to program all three sensors from the NoteLab kit, verifying the results with the \emph{Serial Monitor}.

\noindent\textbf{Task Unit 4 (Wireless network setup):} In this task unit, the previously used single board computers are used to configure their wireless interfaces. To this end, students are instructed on how to program  microcontrollers to receiving an IP address from the network, as illustrated in Fig.~\ref{fig:iotarduino}. To set up WiFi network, each student is given an access to a NETGEAR router, which is pre-configured with default settings. Students are first asked to connect the router to a RPi in a wired fashion, i.e., via an Ethernet cable, in order to change the default settings and be able to setup a network name of choice in ${\textstyle Name(SSID)}$  field to a value  ${\textstyle netwX}$, and the password in  ${\textstyle Password(Network Key)}$ field to a value  ${\textstyle passwordX}$. The value noted as ${\textstyle X}$ will be replaced with a randomly generated number for each of the RPi devices. Following the same approach the network address is to be set to  ${\textstyle 192.168.X.0}$ (e.g.,, 192.168.1.0, 192.168.2.0, ...) with subnet mask 255.255.255.0. Table~\ref{table:wifi} illustrates network configurations on the example of two students working in the lab (each with their own devices).

Let us illustrate in Fig.~\ref{fig:iotarduino} how an RPi 1 is assigned an IP address from  192.168.1.0 network. Student uses the WiFi credentials illustrated in Table~\ref{table:wifi} to connect their microcontrollers to the network and in that way obtain IP addresses from the same network. In this example, it is necessary to include a few Arduino IDE WiFi libraries for ESP8266 boards as shown in the Arduino code block (the libraries included are \emph{WiFiClient.h} and \emph{ESP8266WiFi.h}). After uploading the code, the IoT devices (assigned IP addresses ${\textstyle 192.168.1.y}$ and ${\textstyle 192.168.1.z}$ and the RPi device are all connected to the same WiFi network, where the exact assigned IP addresses of microcontrollers can be verified in the Arduino IDE's \emph{Serial Monitor}. This code is now merged with the sensor and actuator sketches from the previous task (with the example for DHT11 provided by instructors), which in turn finally allows students to send the data measured on sensors over a wireless network.

\begin{table}[h!]
  \caption{Wireless network configurations in IoT context}
  \setlength{\tabcolsep}{3pt}
  \begin{tabular}{|p{53pt}|p{24pt}|p{38pt}|p{54pt}|p{53pt}|}
  \hline
 Group & SSID & Password & Network address & Subnet mask\\ 
\hline
Student1(RPi 1) & netw1 & password1 & 192.168.1.0  & 255.255.255.0\\ 
\hline
Student2(RPi 2) & netw2 & password2 & 192.168.2.0  & 255.255.255.0\\ 
\hline
\end{tabular}
\vspace{-0.2cm}
\label{table:wifi}
\end{table}

\vspace{-0.2cm}
% Edge setup
\subsection{Edge computing context (Task Units 4-7)}\label{edge}
The devices used in edge computing context include primarily single board computers (RPi 3 Model B+ or RPi 4) but also a desktop computer running an edge-cloud connector.  RPis are configured to run MQTT communication protocol software, including MQTT broker and the related processing software. The main purpose of the edge-cloud connector is as the name says to connecting the devices in edge and cloud context. To understand the role of communication protocols, students are given instruction, including the learning material about state of the art application layer protocols that are currently being implemented in IoT resource constrained environments. This group of Task Units also connects IoT and edge computing context, and students are tasked with reading and measuring performance of the data published by sensors in IoT context. As previously noted, each student creates a separate group of edge devices. However, all groups of edge devices stream the data measured to the same edge-cloud connector.  

\noindent\textbf{Task Unit 5 (MQTT broker installation):} The devices in the edge context are here to be setup as MQTT brokers. The broker is responsible for receiving  the data generated in form of MQTT messages, and then publishing the received messages to all subscribed clients. To install the MQTT broker, students use an open source broker Mosquitto, pre-installed as docker image made available in the official repository of container images, Docker Hub. This allows also students to  get introduced to the containerization concept \cite{Knoche}. For pre-installation on RPis, instructors need to download the newest ARM compatible official Eclipse Mosquito docker image. This in turn requires to install Docker software first, with the installation steps notably different for various models of RPis (3 or 4). With Docker installed and running, students run MQTT broker on port 1883 (this port is usually used for MQTT brokers) with a simple command \emph{sudo docker run -p 1883:1883 eclipse-mosquitto}. The output of will be a message indicating that broker is listening for incoming messages on the port 1883. Upon completion of this task unit, the edge devices are ready for microcontrollers to connect to the MQTT broker and send the sensor data from the IoT context.

\begin{figure}[h]
  \centering
  \includegraphics[width=0.91\columnwidth]{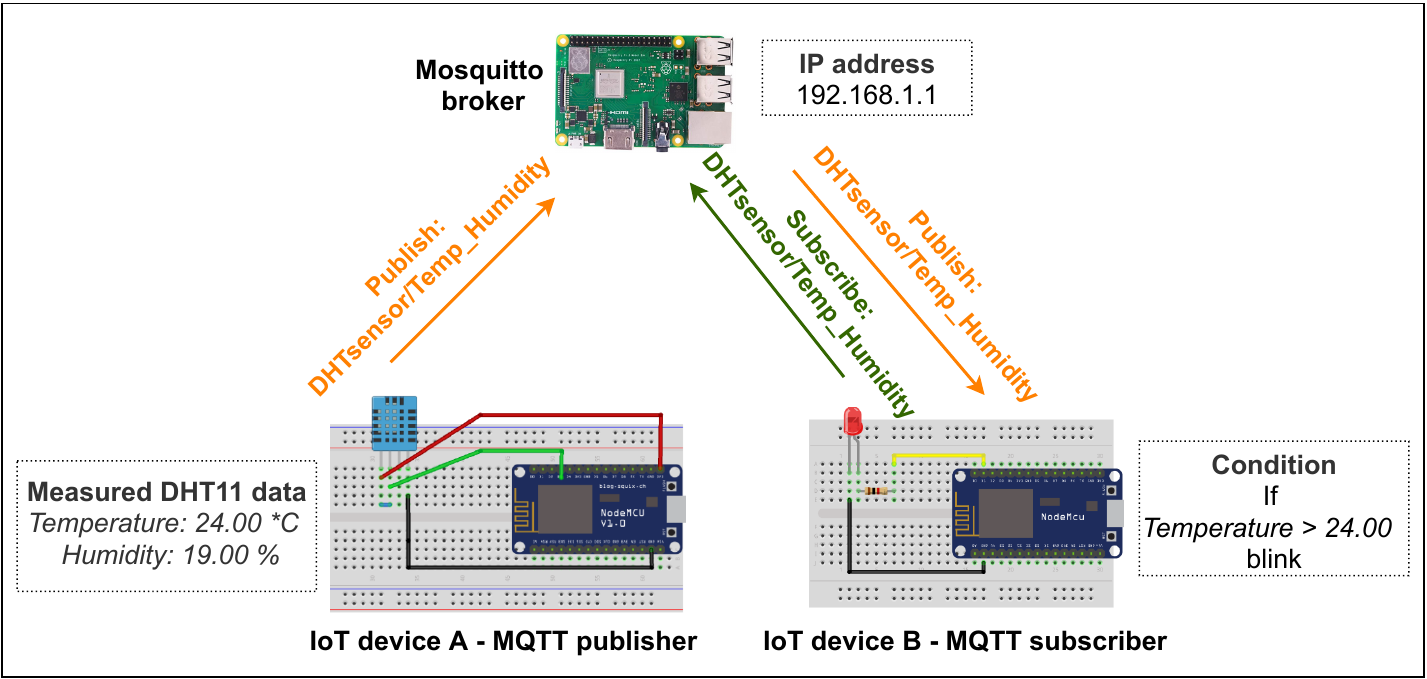}
  \caption{\label{fig:mqttcomm}MQTT communication exchange between IoT and edge context}
  \vspace{-0.3cm}
\end{figure}

\noindent\textbf{Task Unit 6 (MQTT communication):}  In this task unit, students learn how to program MQTT publisher and subscriber clients on microcontrollers and establish the communication with the broker, over the previously set WiFi network from Task Unit 1.4. The subscriber and publisher clients are implemented on two microcontroller boards.  The subscriber is implemented on the board connected to the LED actuator while publisher connects to the sensors. (Recall that the setup of the board circuits was completed in Task Unit 1.2). The MQTT communication exchange between a publisher and a subscriber is shown in Fig.~\ref{fig:mqttcomm}, on the examples of DHT11 sensor and LED actuator, respectively. Publisher, which is implemented on microcontroller connected to DHT11 sensor (IoT device A),  publishes the information based on the sensor's readings, i.e., temperature and humidity values. The data is published to a topic \emph{DHTsensor\textbackslash Temp\textunderscore humidity} and subscriber subscribes to this data. All communications goes through MQTT broker on RPi which listens for MQTT messages on port 1883. For the subscriber - actuator, a condition needs to be added that LED is to be turned on in case the temperature readings from the publisher exceed the threshold values. For example, this can be defined as "turn on the LED if temperature exceeds 22 degrees Celsius"). Based on the wireless network setup (Task Unit 1.3), publisher and subscriber are assigned the IP addresses. In this case, the broker can be reached on IP address of the Raspberry Pi's wireless interface, based on network configuration from Table~\ref{table:wifi}. In the example of the student working on the RPi 1 with SSID and password corresponding to ${\textstyle netw1}$ and ${\textstyle password1}$, IP address of the Raspberry Pi wireless interface, as well as of the MQTT broker will be assigned from the network 192.168.1.0/24. By simply typing \emph{ifconfig} in their Raspberry Pi's terminal they will be able to find out broker's IP (e.g., 192.168.1.1), which is necessary to program publishers and subscribers, as described next. 

\begin{figure*}[h]
  \centering
  \begin{subfigure}[b]{0.453\textwidth}
      \centering
      \includegraphics[width=\textwidth]{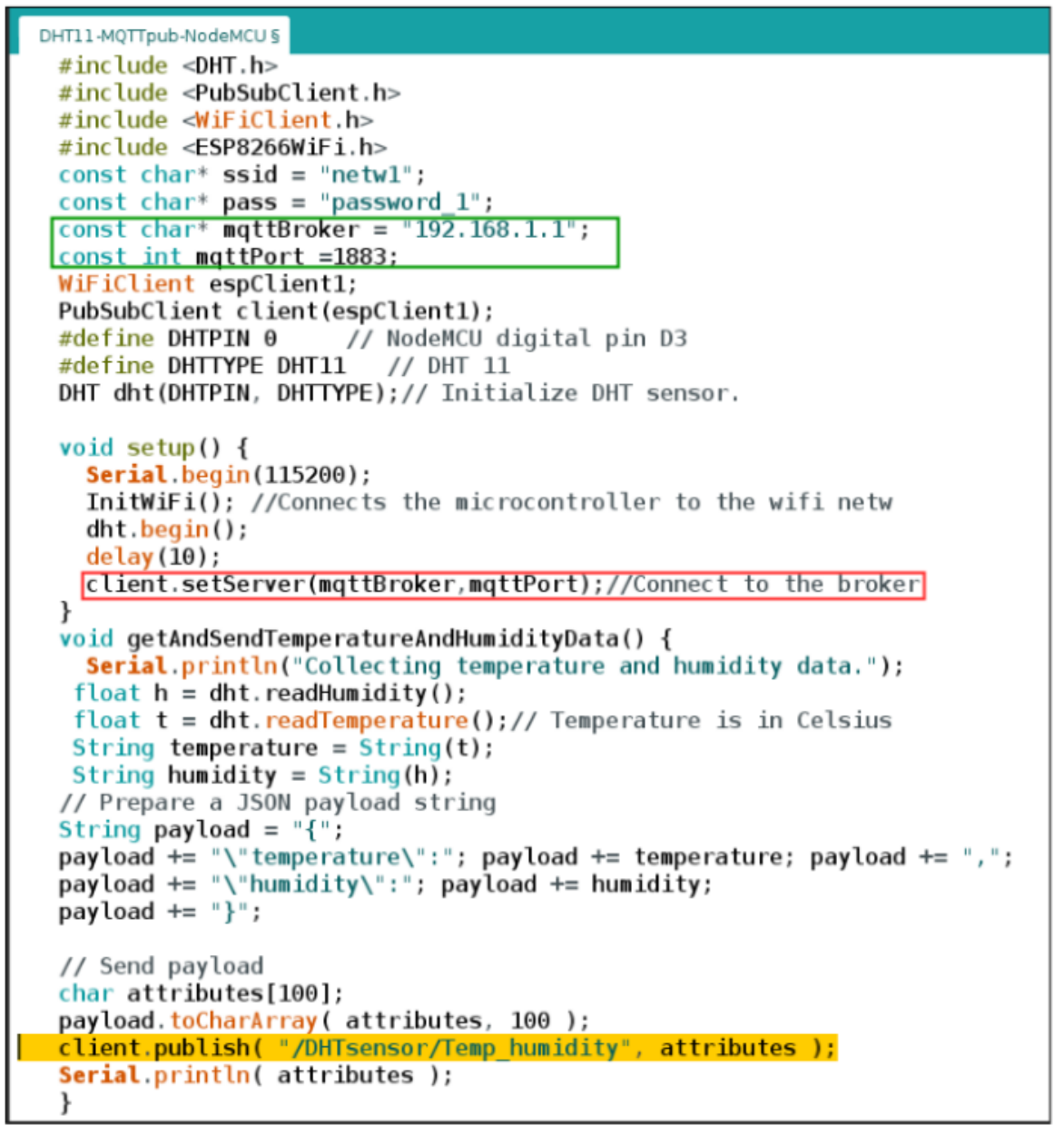}
      \caption{MQTT publisher}
      \label{fig:1a}
  \end{subfigure}
  \hfill
  \begin{subfigure}[b]{0.47\textwidth}
      \centering
      \includegraphics[width=\textwidth]{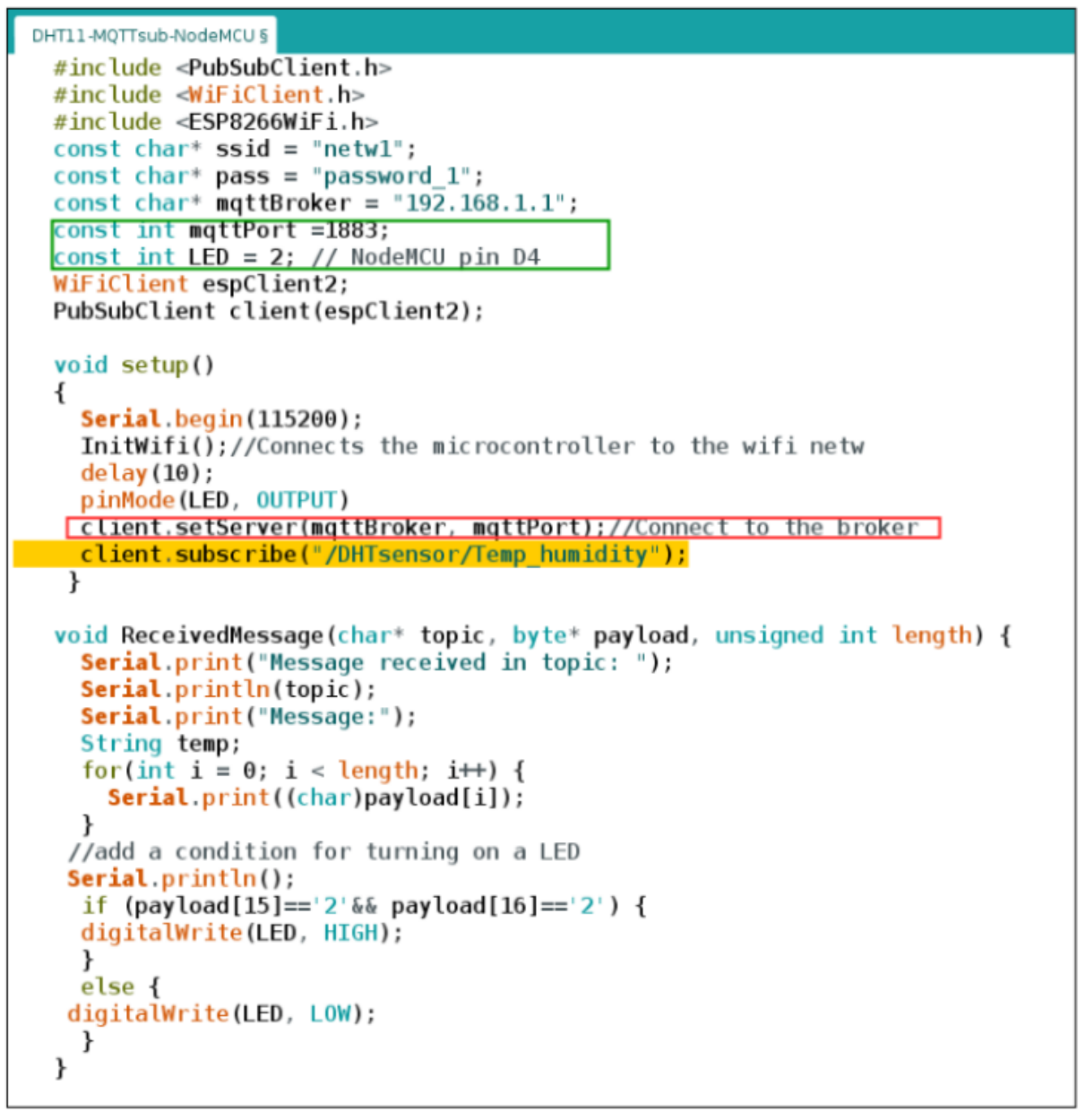}
      \caption{MQTT subscriber}
      \label{fig:1b}
  \end{subfigure}
  \vspace{0.2cm}
     \caption{\label{fig:pubsub} Arduino IDE programming of MQTT communication}    
     \vspace{-0.4cm}    
\end{figure*}

The said programming of MQTT based communication in Fig.~\ref{fig:mqttcomm} is illustrated in Fig. ~\ref{fig:pubsub}.  The main function and parameters for Arduino IDE sketches of MQTT publisher and subscriber in Arduino IDE are shown on the examples of DHT11 sensor and LED actuator, respectively. 
The outputs of running the code for subscriber and publisher are as follows:

\lstset{numberstyle=\tiny, morecomment=[l]{//}, commentstyle=\color[rgb]{0,0,0.70}\ttfamily\footnotesize}  
\begin{lstlisting}[linewidth=\columnwidth,breaklines=true,basicstyle=\footnotesize]
  // Publisher output:
   Wifi connected
   IP address is:
   192.168.1.2
   Connecting to Mosquitto Broker [DONE]
   Humidity: 19.00 %
   Temperature: 22.00 *C
   Sending temperature and humidity : 
   [22.00, 19.00] -> 
   {"temperature":22.00, "humidity":18.00}
   // Subscriber output:
   Wifi connected
   IP address is:
   192.168.1.3
   Connecting to Mosquitto Broker [DONE]
   Message received in topic: 
   DHTsensor/Temp_humidity
   Message: {"temperature":25.00, "humidity":36.00}
  \end{lstlisting}

  \vspace{-0.2cm}

\begin{figure}[h]
    \centering
    \includegraphics[width=0.91\columnwidth]{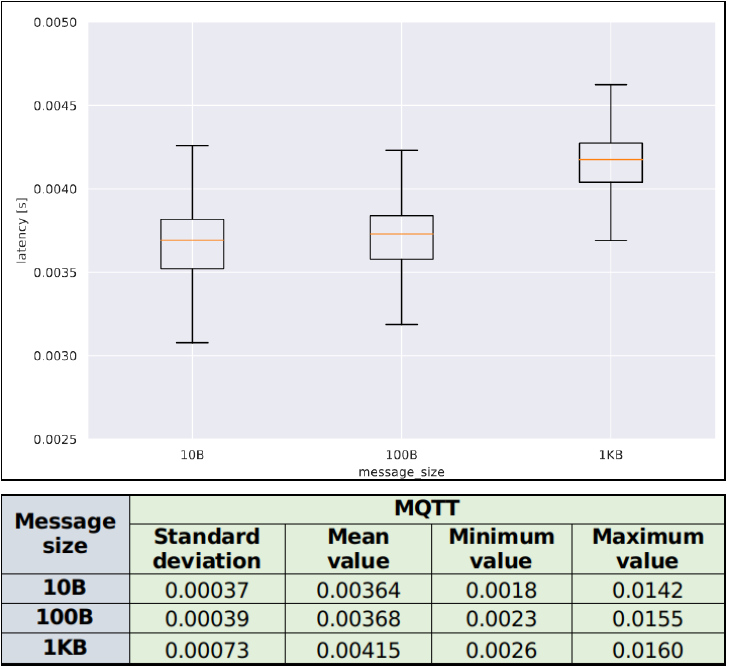}
    \caption{\label{fig:exp1}MQTT latency measurements  in edge context}
    \vspace{-0.6cm}
  \end{figure}

  \noindent\textbf{Task Unit 7 (Latency measurements)} In this task unit, students measure one of the most critical performance indicators that is the communication latency. We start with the publisher creating the message load and creating a timestamp before sending the data to the broker. Student can measure the time for the message to reach the subscriber by creating another timestamp. The time that passes between the two timestamps is the latency value measured. The experiment can be repeated for different message sizes, and the results can be statistically evaluated. Fig.~\ref{fig:exp1} shows statistical results (minimum, maximum, mean values and standard deviation) of one such experiment in form of a table and a boxplot, for message sizes of 10B, 100B, and 1KB. From these measurements, students can learn that the mean value of MQTT latency is around 0.004 sec and that the size of the message did not have a major effect. The boxplot results also show that the latency for 10B message size is in the range of 0.00307s - 0.00425s, for 100B message size 0.00318s - 0.00422s, and for 1KB it is 0.00368s - 0.00462s. This simple statistical analysis introduces students also to research in this field. 

Additional measurement assignments for students that completed the previous task units in shorter time-frame include measuring power consumption for different application layer protocols. This is particularly interesting for the most critical component regarding its resource capabilities, which is the device chosen for the IoT context - microcontroller. Here, students were instructed to estimate ESP8266 microcontroller board RAM utilization for sending the sensor data, by measuring its dynamic memory allocation (free Heap size). This was achieved by sending a fixed number of messages containing sensor measured data in JSON format (i.e 1000 temperature values) to the Mosquito broker and returning as the output the free heap size in bytes. To ensure a correct estimation several measurements were taken and the mean value calculated. The experiment was then repeated with the HTTP protocol implementation.
The obtained values in table \ref{lab:heap} indicate that the free heap size is lower for HTTP, which means that for sending the same number of messages containing sensor data, using HTTP will result in microcontroller's higher RAM utilization compared to the MQTT. 

\begin{table}[!htbp]
  \caption{Microcontroller RAM utilization for sending sensor data to the edge}
\centering
\begin{tabular}{|c|c|c|c|}
\hline
protocol & Free Heap in Bytes \\ \hline
MQTT     & 49142              \\ \hline
HTTP     & 48932              \\ \hline
\end{tabular}
\label{lab:heap}
\vspace{-0.3cm}
\end{table}

\subsection{Cloud computing context}\label{cloud}
The cloud computing context requires the implementation of Kafka based edge-cloud connector that connects devices in edge computing context to Firebase cloud. To this end, we pre-install the desktop computer in edge context to run a dockerized \emph{Kafka-Firebase connector} solution to all edge devices. To establish the IoT-edge-cloud communication, all edge devices in edge context need to be in the same network. We setup static IP addresses for Ethernet interfaces on each devices in edge context, as illustrated in Table~\ref{table:wired}. As the connector is not only shared by all RPi devices but also by all students controlling their corresponding RPis, the instructors manually assign static IP address on two Ethernet interfaces. One address is used for the network configured to connect to the cloud database, while the other for RPi used in the edge context, see Table~\ref{table:wired}. When all of the IP addresses have been configured students will by using the Ethernet cables connect their nodes to a switch, with one port connected with a PC, now having their entire edge domain in the same network.

\begin{table}[h!]
  \caption{Configuration of static IP addresses}
  \setlength{\tabcolsep}{3pt}
  \begin{tabular}{|p{128pt}|p{54pt}|p{53pt}|}
  \hline
 Edge context & IP address & Subnet mask\\ 
\hline
Desktop PC (edge-cloud connector) & 192.168.5.2  & 255.255.255.0\\
\hline
Student1 (RPi 1) & 192.168.5.3 & 255.255.255.0\\ 
\hline
Student2 (RPi 2) & 192.168.5.4  & 255.255.255.0\\ 
\hline
\end{tabular}
\label{table:wired}
\vspace{-0.2cm}
\end{table}

\noindent\textbf{Task Unit 8 (Configuring cloud database):} To get started, students first sign in on the edge-cloud connector device with Google account created for NoteLab. This account allows the student to use Firebase Console (\emph{https://firebase.google.com}) and create a new \emph{Firebase project.} %One such project named NOTElabsummer2020 is illustrated in Fig.~\ref{fig:firebase1}.
After creating a project, students select from multiple database options, and by doing so are required to opt for a real-time database. The database can be started in two modes, the so-called locked and test mode. In NoteLab, we use the test mode (the locked mode is for production solutions). Whenever a new database is created, it will have a unique URL ending in \emph{firebaseio.com} and the URL of the database following the format \emph{\textless project \textunderscore id\textgreater .firebaseio.com}.% In the example shown, the URL of the NoteLab database corresponds to: https:\slash \slash notelabsummer2020-b4086.firebaseio.com.
 This information is later used for interfacing the connector and Firebase database. In this stage, students have an empty database in Firebase cloud. The last pair of information required to connecting the connector in the edge context to the database in the cloud context are the host name and authorization key/secret key of the created project. This information can be read from \emph{Project Settings-Service accounts} tab where the option to generate a new private key will be offered for generation. Students can save the key as \emph{firebase-admin.json} in the connector.

\noindent\textbf{Task Unit 9 (Connecting edge and cloud databases):} Since this is an undergraduate course, where students are not required to learn all the development details in Java, the connector device is pre-configured, with all relevant details  saved locally, including the installation of Docker. Furthermore, the docker-compose tool is also pre-installed allowing the deployment and management of multiple containers at the same time instead of running each container individually. This is due to the fact that the Kafka based connector (\emph{Kafka-Firebase connector} from Fig.~\ref{fig:arch}b) actually includes multiple containers. Based on pre-installation, students receive the set of instructions, which consist of modifying two parameters found in \emph{docker-compose.yml} saved locally. File \emph{docker-compose.yml} can be opened in a text editor and two parameters need to be configured: ${\textstyle external IP address}$ and ${\textstyle FIREBASE\textunderscore URL}$. This way, students have configured the database for synchronization between edge and cloud context.

\begin{figure}[h]
  \centering
  \includegraphics[width=0.91\columnwidth]{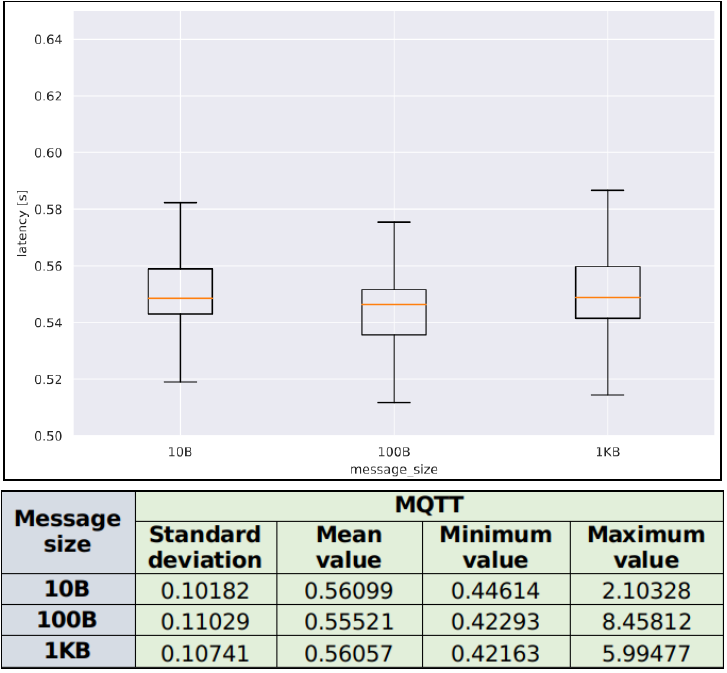}
  \caption{\label{fig:exp2}MQTT latency measurements IoT-E-C}
  \vspace{-0.4cm}
\end{figure}

\noindent\textbf{Task Unit 10 (IoT-E-C performance measurements):} We illustrate end-to-end latency measurement similar to the IoT-to-edge measurements from Task Unit 7, where latency was measured as the time that passes between the two timestamps. Here, the first timestamp is also created with a MQTT publisher sending the message to the broker. The second timestamp is created when the message previously sent is received by the subscriber directly from the Firebase which represents the cloud context. The experiment is again repeated for different message sizes. Fig. ~\ref{fig:exp2} shows statistical results in form of a table and a boxplot, for different message sizes of 10B, 100B, and 1KB. Students can notice that compared with Fig. \ref{fig:exp1} the end-to-end latency values were noticeably higher than what was measured between IoT devices and edge.

\subsection{Lessons learnt}\label{lessons}

Undergraduate students typically come with different knowledge backgrounds and different programming and electronics class prerequisites. At the same time, most students already have specific preferences towards either software- or hardware-oriented courses. Students with knowledge in programming and software engineering performed especially well in Task Units related to data processing and storage in IoT-E-C context (Task Units 7, 8, 9). Students more interested in electronics aspects of the course, usually excel at Task Units 1, 2, and 3. Clearly, the benefit of the course is that a number of task units is designed to combine various backgrounds and preferences (Task Units 4, 5, 6, and 10). 

To motivate students without a certain background, it was essential to create a detailed instruction guidelines that cover a wide range of topics, from devices' hardware capabilities, circuit setup schematics over to programming scripts and basic instructions in various programming languages. Even then students faced challenges. For instance,  when configuring hardware components, and despite following a precise circuit schematics,  minor omissions in interfacing sensors and actuators inevitably happen; wrongly connected jumper wire directly leads to device damage. This was a common occurrence,  and we found that it helped developing student's engineering and problem solving skills. On the flip side, the course needs to purchase backup hardware components.

In terms of developing and testing software solutions, a set of basic and detailed instructions for programming in Arduino IDE is rather critical, since also the students versed in programming may have no knowledge in low-level programming languages. While learning the programming basics with Arduino IDE is not complex in itself, students need to get used to establishing a rather strong link between software and hardware components in IoT context. In the end, students can learn that even a minor differences in hardware configurations would end up in the code being executed differently for different devices (and different students in charge of these devices). Anecdotally, one of the common difficulties for students is to differentiate whether the device is damaged or simply disconnected, as in both cases the low-level program would just stop running. 

With dockerized components, on the other hand, students can face the challenge on how to run docker images even when already prepared for them (including the instructions related to operating systems, e.g., Ubuntu and Raspbian OS). For the development of their own application scripts, we found that it was advantageous to students to have basic knowledge in Python (Python 2 and Python 3 come pre-installed on Raspbian OS). The Python programming skills are not mandatory though, as the only scripted application developed in Python was mqtt proxy. As both RPi and Desktop PC operating systems are Debian-based, the students are instructed to use \emph{commandline} interface instead of graphical user interface for which also basic skills need to be acquired.%These are the issues that students will often not be able to solve on their own and have to be particularly taken into consideration by instructors.

All challenges that students face require help of tutors and instructors. In addition, instructors also face conceptual challenges. One of the daunting challenges is to scope the learning framework, given the rapid evolution of the state-of-the-art, both in hardware and software. Instructors also need to provide a significant amount of up-to-date, detailed and workable instructions to students about the devices' hardware and software capabilities as well as the networking and communication protocols. Table~\ref{table:instr} summarizes hardware and software ready-configurations and code that instructors need to prepare. As it can be seen, scoping the learning framework requires not only students to further challenge their hardware and software skills, but also instructors to provide the enabling hardware and software configurations. 

\begin{table}[h!]
  \caption{Material provided by instructors}
  \setlength{\tabcolsep}{3pt}
  \begin{tabular}{|p{30pt}|p{87pt}|p{116pt}|}
  \hline
 Context &  Hardware and OS &  Software\\ 
\hline
\textbf{\vfill IoT}&
\begin{itemize}[leftmargin=*]
  \vspace{-0.2cm}
  \item Pin layout, specifications of sensors, actuators and microcontrollers
  \item Breadboard circuit scheme for connecting sensor/actuators with microcontroller
 \vspace{-0.2cm}
\end{itemize}
& 
\begin{itemize}[leftmargin=*]
  \vspace{-0.2cm}
  \item Arduino IDE configured for ESP8266 
  \item List of libraries to program sensors and actuators
  \item Arduino sketch code: sensors,  actuators, wireless network connection, MQTT publisher/subscriber.
  \vspace{-0.2cm}
 \end{itemize}
\\
\hline
\textbf{\vfill Edge}& 
\begin{itemize}[leftmargin=*]
  \vspace{-0.2cm}
  \item RPi 3/4 Model B pin layout and specifications
  \item Latest version of Raspbian OS 
  \item Breadboard circuit scheme for connecting sensor/actuators with microcontroller
  \item Latest Ubuntu version 
  \vspace{-0.2cm}
\end{itemize}
  & 
  \begin{itemize}[leftmargin=*]
    \vspace{-0.2cm}
    \item Mosquitto broker docker image downloaded with installation and configuration instructions 
    \item Instructions on RPi and Desktop PC static IP address configuration
    \item mqtt proxy python script
    \item Edge-cloud connector docker image downloaded  with installation and configuration instructions
    \vspace{-0.2cm}
  \end{itemize}
  \\
  \hline
  \textbf{\vfill Cloud} & 
  & 
  \begin{itemize}[leftmargin=*]
    \vspace{-0.2cm}
    \item NoteLab Google account to sign in to Firebase
    \item New Firebase project and Real-time Database created 
    \item Kafka based edge-cloud connector connected to the database
    \vspace{-0.2cm}
  \end{itemize}\\
  \hline
\end{tabular}
\label{table:instr}
\vspace{-0.5cm}
\end{table}

\section{Conclusions and Outlook}\label{conclusion}
We proposed and outlined the design and development of an undergraduate laboratory course, we named Network-of-Things Engineering Lab (NoteLab), - that set the goal of implementing various interfaces and communication protocols to connect IoT, edge and cloud computing systems and evaluate their performance. Unlike other such related courses, our course was designed to provide an implementation of the entire system, based on open-source and low-cost devices, and for the subsequent evaluation and benchmarking of the performance of the entire system. We also integrated various communication protocol and interface solutions.   

In this paper, we focused on technical aspects of the course implementation. Despite receiving overwhelmingly positive feedback from students, we have not yet didactically and competently analyzed the learning outcomes and student assessments. This is also due to the course duration (one semester over two years) and due to the limited number of students per lab (six). 
In the future, we plan to leverage two computer science related learning taxonomies, Bloom and SOLO taxonomy, which will allow us to improve upon design of the course didactically. Future possible modular extension to our course include other application layer protocols, most notably HTTP3, security protocols, and machine learning applications.

\bibliographystyle{IEEEtran}  

\bibliography{notelabbibliography}

\end{document}